\documentclass[%
 reprint,
superscriptaddress,
 amsmath,amssymb,
aip,
jcp,
]{revtex4-1}

\usepackage{amsmath}
\usepackage{amssymb}
\usepackage{amsfonts}
\usepackage{graphicx}
\usepackage{subfigure}
\usepackage[version=3]{mhchem} 
\usepackage{multirow}
\usepackage{tabularx}
\usepackage{booktabs}
\usepackage{ctable}
\newcolumntype{Y}{>{\raggedright\arraybackslash}X}      
\newcolumntype{W}{>{\raggedleft\arraybackslash}X}       
\newcolumntype{Z}{>{\centering\arraybackslash}X}        

\usepackage{dcolumn}

\usepackage{lmodern}        
\usepackage[T1]{fontenc}
\usepackage{microtype}

\begin{document}

\title{Liquid water through density-functional molecular dynamics: Plane-wave vs.\ atomic-orbital basis sets}

\author{Giacomo Miceli}
\email{giacomo.miceli@epfl.ch}
\affiliation{Chaire de Simulation \`a l'Echelle Atomique (CSEA), %
 Ecole Polytechnique F\'ed\'erale de Lausanne (EPFL), %
 CH-1015 Lausanne, Switzerland}
 
\author{J\"urg Hutter}
\affiliation{Department of Chemistry, University of Z\"urich,
 Winterthurerstrasse 190, CH-8057 Z\"urich, Switzerland}

\author{Alfredo Pasquarello}
\affiliation{Chaire de Simulation \`a l'Echelle Atomique (CSEA), %
 Ecole Polytechnique F\'ed\'erale de Lausanne (EPFL), %
 CH-1015 Lausanne, Switzerland}
 
\date{\today}

\begin{abstract}
We determine and compare structural, dynamical, and electronic properties of liquid
water at near ambient conditions through density-functional molecular dynamics 
simulations, when using either plane-wave or atomic-orbital basis sets. In both 
frameworks, the electronic structure and the atomic forces are self-consistently
determined within the same theoretical scheme based on a nonlocal density functional
accounting for van der Waals interactions. The overall properties of liquid water
achieved within the two frameworks are in excellent agreement with each other. 
Thus, our study supports that implementations with plane-wave or atomic-orbital 
basis sets yield equivalent results and can be used indiscriminately in study of
liquid water or aqueous solutions. 
\end{abstract}

\maketitle

\section{Introduction}
\label{sec:intro}
Liquid water plays a fundamental role in a multitude of phenomena of primary 
relevance to diverse areas of science. It is thus not surprising that in the past
30 years many efforts have been invested in better understanding liquid water
properties at the molecular level. In this context, molecular dynamics and Monte
Carlo simulation techniques have been largely employed as a complementary tool 
to experiment to investigate the nature of water at the atomic scale. Indeed, the
increasing availability of computer resources and the improvement of computational
algorithms have resulted in an accurate description of intermolecular interactions
by the direct evaluation of the evolving electronic structure. In this respect, 
the Car-Parrinello method\cite{car_PRL1985} has been instrumental to unify density
functional calculations and molecular dynamics simulations in the study of liquid
water.\cite{laasonen_JCP1993,sprik_JCP1996} In fact, {\it ab initio} molecular 
dynamics simulations represent an invaluable tool to simultaneously access its 
structural, dynamical, \cite{laasonen_JCP1993,%
sprik_JCP1996,todorova_JPCB2006,lee_JCP2007,schmidt_JPCB2009,distasio_JCP2014,%
miceli_JCP2015,delben_JCP2015} and electronic \cite{silvestrelli_JCP1999,%
prendergast_JCP2005,ambrosio_JCP2015,chen_2016} properties.

Most of the first-principles simulations of liquid water have been performed 
within the theoretical framework of density functional theory with a (semi)local 
approximation for the exchange-correlation energy.\cite{laasonen_JCP1993,sprik_JCP1996,
lee_JCP2007} However, it has become clear that these approximations lead to a poor
description of the structural and the dynamical properties. Liquid water is found
to be overstructured, shows a very low diffusion coefficient, and its equilibrium mass
density underestimates the experimental value by about 15\%.\cite{schmidt_JPCB2009,%
wang_JCP2011,miceli_JCP2015} These shortcomings still persist when the electronic
structure is described at a higher level of theory, such as with hybrid density 
functionals. \cite{todorova_JPCB2006,delben_JPCL2013,delben_JCP2015,gaiduk_JPCL2015,ambrosio_2016} Similarly, 
from the more fundamental side, a quantum treatment of the nuclei slightly modifies
the properties of the liquid. \cite{morrone_PRL2008,giberti_JPCB2014,delben_JCP2015}
A substantial improvement is instead achieved when van der Waals interactions are
correctly accounted for.\cite{schmidt_JPCB2009,lin_JPCB2009,wang_JCP2011,lin_JCTC2012,%
delben_JPCL2013,distasio_JCP2014,miceli_JCP2015,gaiduk_JPCL2015,delben_JCP2015,ikeda_JCP2015}
For a broader overview on the performance of various popular exchange-correlation
functionals in simulating liquid water, we refer the readers to the recent
review by Gillan, Alf\`e, and Michaelides.\cite{gillan_JCP2016}

The need to perform accurate simulations of liquid water has brought the attention
to post Hartree-Fock methods. Indeed, simulations of the liquid using sophisticated
and accurate electronic-structure methods, such as the second-order M{\o}ller-Plesset
approximation (MP2)\cite{delben_JPCL2013} and variational quantum Monte Carlo%
\cite{zen_JCP2015} have been already reported in the literature. However, in spite
of their accuracy, these methods are still computationally too demanding for a 
widespread use in routine simulations.

More recently, the need to accurately describe the electronic structure of liquid
water has been constantly increasing. Indeed, this represents a major prerequisite
for further progress in the design of new and efficient systems for photocatalytic
water splitting.\cite{gratzel_nature2001,walter_CR2010,adriaanse_JPCL2012,ambrosio_JCP2015}
To this aim, it has become imperative to resort to fully {\it ab initio} schemes
which can shed light onto the electronic structure of the liquid without the use of
any phenomenological parameters. In this respect, it has recently been shown that
the combination of path-integral molecular dynamics simulations and quasiparticle
self-consistent {\it GW} accounting for vertex corrections correctly reproduces 
the experimental photoemission spectrum.\cite{chen_2016}

It has become clear that a high level of theory is needed for an accurate 
description of liquid water. The scientific efforts in the last years have 
led to significant improvements of the available computer codes, which now allow 
for highly accurate simulations of liquid water. 
{\tt CP2K} is one of the most versatile suite of programs used to perform atomistic
simulations of solid-state, liquid, and biological systems.\cite{cp2k} In {\tt CP2K}, quantum
chemistry simulations are performed in the framework of density functional theory
relying on a mixed scheme based on Gaussian and plane-wave basis sets. Within this
scheme, a wide variety of approximations for the exchange-correlation functional
are implemented. The efficient use of basis sets and the availability of advanced
algorithms, make of {\tt CP2K} one of the most suitable and cost-effective codes
for carrying out large scale first-principles simulations. Another largely used 
software suite to perform quantum chemistry and material science simulations is 
{\sc Q}uantum-{\sc Espresso}.\cite{quantum_espresso} {\sc Q}uantum-{\sc Espresso}
has specifically been designed
to perform highly accurate electronic structure calculations. This accuracy rests
on the use of plane-wave basis sets in conjunction with pseudopotentials. An important 
advantage of this scheme is the possibility of verifying the completeness of 
the adopted basis set by the adjustment of a single parameter corresponding to 
the kinetic energy cutoff of the plane waves. Furthermore, as shown in a recent 
community study, currently implemented pseudopotentials offer a high degree of
accuracy and reproducibility with respect to all-electron calculations.%
\cite{lejaeghere_science2016}

Most simulations on liquid water are currently performed with either atomic-orbital
based codes such as {\tt CP2K} or with plane-wave-based codes such as 
{\sc Q}uantum-{\sc Espresso}. In doing so, the simulation parameters used in the
two codes are set primarily on the basis of energy convergence criteria as they
can be tested separately in the two codes. It is generally assumed that the adopted 
protocols lead to similar descriptions of the structural properties of liquid water,%
\cite{kuo_JPCB2004}
but the comparison for the electronic and dynamical properties is not as well documented. 
Nevertheless, the agreement between different protocols is crucial for the general 
development of this research area and would indicate that the calculated properties 
are converged. In this respect, it is critical to compare simulations performed with 
identical exchange-correlation functionals and to adopt standard simulation parameters 
as currently in use in the research community. In particular, the use of the 
Car-Parrinello method is inappropriate for such a comparison as it could be biased 
by the use of a fictitious mass. Indeed, the use of fictitious masses are known 
to affect the detailed MD trajectories\cite{grossman_JCP2003,tangney_JCP2002} and
to be dependent on the adopted basis set.\cite{car_PRL1985}

In this work, we compare the results of first-principles molecular dynamics simulations
of liquid water performed with the {\tt CP2K} and the {\sc Q}uantum-{\sc Espresso}
codes, adopting standard simulation protocols separately set for the two codes.
The molecular dynamics simulations in both the NVE and NpH ensembles are carried out 
within the Born-Oppenheimer approximation which does not present limitations that 
could be biased by the adopted basis set. In both cases, we use the same nonlocal 
exchange-correlation density functional which explicitly accounts for van der Waals
interactions. Our results show that the two approaches are in excellent agreement 
with each other leading to an equivalent description of the structural, dynamical,
and electronic properties of liquid water.

\section{Computational details}
\label{sec:methods}
Throughout this work, the electronic structure and the atomic forces are calculated
within the self-consistent Kohn-Sham approach to density functional theory (DFT). We 
perform simulations in which van der Waals interactions are explicitly taken into
account. To this aim, among the several theoretical schemes proposed, we adopt the
nonlocal rVV10 van der Waals functional.\cite{sabatini_PRB2013} This is essentially
a revised but equivalent formulation of the VV10 nonlocal functional recently 
introduced by Vydrov and Van Voorhis.\cite{vydrov_JCP2010}
Within this scheme, we use a semilocal exchange-correlation functional which
results from a combination of the refitted Perdew-Wang exchange 
functional\cite{murray_JCTC2009} and the local density approximation to the 
correlation according to the Perdew-Wang parametrization.\cite{perdew_PRB1992}	
This semilocal exchange-correlation functional is then augmented with a nonlocal 
part which accounts for dispersion interactions. The result is a very simple analytic 
form which depends on an empirically determined parameter $b$, which controls the 
short-range behavior of the functional.\cite{vydrov_JCP2010,sabatini_PRB2013}
This functional can be used to reproduce the correct physical properties of 
weakly bonded systems after an appropriate optimization of the $b$ parameter.
For instance, the best description of the $S22$ set of molecules has been obtained 
for $b=6.3$.\cite{sabatini_PRB2013}
However, while the binding energy and the geometry of weakly bonded molecules 
are well described, rVV10 with this $b$ parameter has been shown to overestimate 
the binding energy of the layered solids.\cite{bjorkman_PRL2012} Bj\"orkman has 
demonstrated that an improved description could be achieved by 
setting the $b$ parameter to higher values (up to 10.25).\cite{bjorkman_PRB2012}
More recently, Miceli, de Gironcoli, and Pasquarello have shown that the rVV10 
functional with the $b$ parameter set to 9.3 (rVV10-b9.3) yields structural 
properties of liquid water at near ambient conditions in very good agreement with
experimental data.\cite{miceli_JCP2015}

\begin{figure}
\centering
\includegraphics[width=8.5cm]{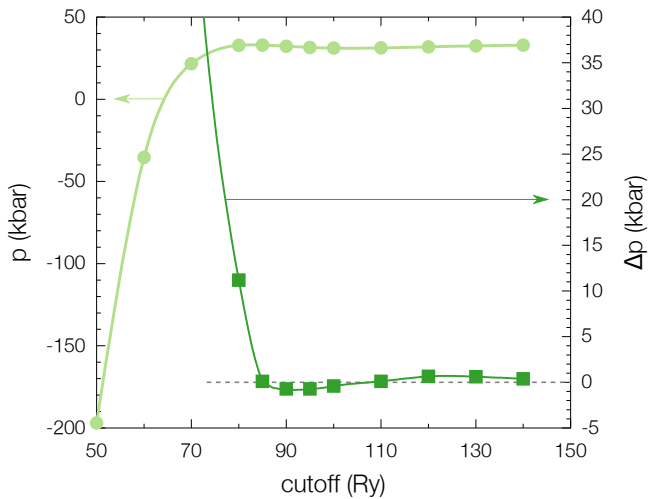}%
\caption{Pressure as a function of the plane-wave energy cutoff for a given structural 
         configuration of liquid water (vertical axis on the left). The second 
         curve represents the difference between the current pressure and that 
         at the previous energy cutoff (vertical axis on the right). The continuous 
         lines are guides to the eye.}
 	\label{fig:pressure}
\end{figure}

The theoretical framework described so far is common to our simulations performed 
with the two codes: {\sc Q}uantum-{\sc Espresso}\cite{quantum_espresso} and 
{\tt CP2K}.\cite{cp2k} The fundamental difference between the codes rests on 
the use of different basis sets for the representation of the electron wavefunction.
In {\sc Q}uantum-{\sc Espresso}, the Kohn-Sham orbitals are expanded on plane waves
(PW), while core-valence interactions are described by Troullier-Martins 
norm-conserving pseudopotentials.\cite{troullier_PRB1991}
The PW energy cutoff is set on the basis of convergence tests for the calculated
pressure associated to specific liquid water snapshots. As shown in Fig.\ \ref{fig:pressure},
we verified that by setting the PW energy cutoff at 85 Ry, the pressure is converged
within 1 kbar.\cite{miceli_JCP2015}
In the mixed Gaussian-plane-wave scheme (GPW) in {\tt CP2K}, the core-valence 
interactions are described by Goedecker-Teter-Hutter pseudopotentials.\cite{goedecker_PRB1996}
The Kohn-Sham orbitals are expanded on a Gaussian basis set,\cite{lippert_MP1997} while an 
auxiliary plane waves basis set defined by a cutoff of 800 Ry is used to expand 
the electron density. We here use the triple-$\zeta$ basis set augmented with 
polarization functions (TZV2P).\cite{vandevondele_JCP2007} This basis set has 
been validated through BLYP simulations of liquid water.\cite{vandevondele_JCP2005}  

We perform Born-Oppenheimer molecular dynamics simulations of liquid water within
both the microcanonical (NVE) and the isobaric-isoenthalpic (NpH) statistical 
ensembles. The Born-Oppenheimer scheme is here preferred to the Car-Parrinello 
one as it avoids any dependence on fictitious parameters which might affect the 
MD trajectory\cite{tangney_JCP2002,grossman_JCP2003} and be basis-set dependent. 

The system is modeled using supercells with 64 water molecules subject to periodic
boundary conditions. In this work, we neglect nuclear quantum effects. Newton's 
equations of motion are integrated with a time step of 0.48 fs to correctly sample
the highest frequency of the O-H stretching mode. The energy convergence threshold
for selfconsistency at each Born-Oppenheimer MD steps is set to $5 \times 10^{-11}$
a.u./atom. In the case of NVE simulations, the conservation of total energies is
better than 1 part over 10$^{5}$ over simulation periods of 10 ps, for both the 
PW and GPW schemes.  

The volume of the cell in NVE simulations is fixed at a value corresponding to the
experimental water density. On the contrary, in the NpH simulations the cell volume
is allowed to fluctuate, but is constrained to preserve the initial cell shape of 
cubic symmetry. A vanishing external pressure is imposed through the use of a 
Parrinello-Rahman barostat.\cite{parrinello_PRL1980} The statistical analysis is
performed on runs at 350 K with duration of about 30 ps each. The latter are 
preceded by equilibrium runs of 5 to 10 ps.

When using isobaric first-principles molecular dynamics with a constant number 
of plane waves, the fluctuations of the cell volume imply fluctuations of the 
effective energy cutoff defining the basis set. Plane-wave basis sets are used 
by both {\sc Q}uantum-{\sc Espresso} and {\tt CP2K}. For this reason, further 
precautions are needed. In {\tt CP2K}, a reference cell with a larger volume is 
used to determine the number of grid points which is then kept fixed regardless 
of the actual size of the simulation cell.\cite{mcgrath_CPC2005,schmidt_JPCB2009}
In {\sc Q}uantum-{\sc Espresso}, the MD runs are restarted when the density of 
the simulation cell reaches excessively low values. The restart at a larger initial
volume restores the energy cutoff required for achieving converged results.%
\cite{miceli_JCP2015}   

For information, we here report the computational performance of the two simulation 
set-ups as recorded for runs on a Cray XC30 system. Using the same number of 
processors, we find that the wall-clock time for one MD step is 21.4 s step for 
{\sc Q}uantum-{\sc Espresso} and 10.3 s for {\tt CP2K}. We record linear scaling
for both simulation schemes up to 128 cores, with efficiencies of 85\% and 90\% 
for {\sc Q}uantum-{\sc Espresso} and {\tt CP2K}, respectively.

\section{Results and discussion}
In this section, we compare and discuss the results of structural, dynamical, and
electronic properties of liquid water at near ambient conditions obtained using
the {\sc Q}uantum-{\sc Espresso} and {\tt CP2K} suites of programs. When available, 
experimental data are reported for comparison. 


Isobaric molecular dynamics simulations allow for fluctuations of the system 
volume. In particular, the volume reaches the equilibrium value at the hydrostatic
pressure set externally. As already pointed out previously,\cite{schmidt_JPCB2009,%
miceli_JCP2015} first principles methods are able to describe the correct density
of liquid water only when the adopted theory explicitly accounts for van der Waals
interactions. In particular, using {\sc Q}uantum-{\sc Espresso} we have shown that
the nonlocal rVV10 functional yields an equilibrium water density of 0.99 g/cm$^3$
when the phenomenological parameter $b$ is set to 9.3.\cite{miceli_JCP2015} By 
carrying out simulations with the same functional but describing the electronic 
wave functions and density with the mixed Gaussian-plane-wave scheme implemented
in {\tt CP2K}, we find that liquid water shows a density of 1.01 g/cm$^3$, at the
same thermodynamic conditions of 350 K and zero pressure. Using the blocking 
analysis method,\cite{flyvbjergJCP1989} we obtain statistical errors of about 
0.01 g/cm$^3$ in both cases, indicating that the two codes are in excellent 
agreement.

\begin{figure}
\centering
\includegraphics[width=8.5cm]{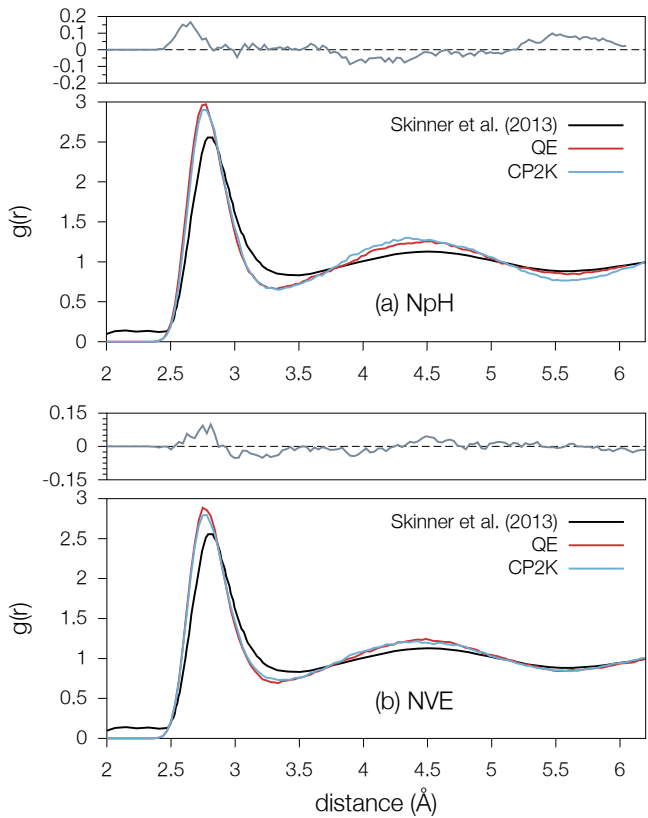}%
\caption{Oxygen-oxygen radial distribution functions resulting from 
         from molecular dynamics trajectories obtained with {\sc Q}uantum-{\sc Espresso}
         (QE) and {\tt CP2K}. For comparison, we superimpose results from the most
         recent experimental work of Ref.\ \citenum{skinner_JCP2013}. Panels (a)
         and (b) refer to isobaric-isoenthalpic and microcanonical simulations, 
         respectively. In both cases, the differences 
         $g_{\textrm{{\sc QE}}}(r) - g_{\textrm{\tt CP2K}}(r)$
         are reported at the top of the respective panels.}
 	\label{fig:gofr}
\end{figure}

To gain deeper insight into the structural properties of liquid water, 
we compare the two calculated oxygen-oxygen radial distribution functions, 
$g_\textrm{OO}(r)$. In Fig.\ \ref{fig:gofr}(a), we show radial distribution
functions as obtained through an isobaric-isoenthalpic sampling with the two different
codes. The difference $g^{\rm NpH}_{\textrm{{\sc QE}}}(r) - g^{\rm NpH}_{\textrm{\tt CP2K}}(r)$
is reported at the top of panel (a) in the same figure. The experimental curve 
from Ref.\ \citenum{skinner_JCP2013} is superimposed. As one can notice, the two
calculated $g_\textrm{OO}(r)$ are in very good agreement. In both cases the 
positions of the first peak are found at 2.75 \AA. The first peak representing 
the first coordination shell is barely higher in the {\sc Q}uantum-{\sc Espresso}
simulation. The same kind of agreement is found for the second shell. 
Compared to experiment, the position of the first peak in the calculated oxygen-oxygen 
radial distribution function is slightly shifted by 0.05 \AA\ towards shorter distances 
with respect to the experimental value. The simulated liquid still shows overstructuring 
with respect to experiment, but the overall description of the system is improved with the 
respect to the PBE functional, as already pointed out previously.\cite{miceli_JCP2015}

We also study the structural properties of liquid water obtained in the NVE
statistical ensemble. We carry out fixed-volume simulations at the experimental 
density of 1 g/cm$^3$ using  {\sc Q}uantum-{\sc Espresso}  and {\tt CP2K} at the
rVV10-b9.3 level of theory. In Fig.\ \ref{fig:gofr}(b), we report the calculated
oxygen-oxygen radial
distribution functions as obtained from these NVE-MD runs. The two calculated
$g_\textrm{OO}(r)$ basically coincide.  {\sc Q}uantum-{\sc Espresso}  gives rise
to an imperceptible overstructuring with respect to {\tt CP2K} at short distances.
At longer distances the agreement between the two codes is excellent. More generally,
we also observe an excellent agreement between the two codes in reproducing the 
overall structural properties of the system. Hereafter, all the presented results
refer to the microcanonical simulations.

Earlier NVE simulations based on the BLYP functional\cite{kuo_JPCB2004} also showed
very good agreement between plane-wave and atomic-orbital schemes in describing the 
structural properties of liquid water. The position and the height of the first peak 
in the $g_{\rm OO}$ differed by only 0.01 \AA\ and 0.2, to be compared with our 
differences of 0.00 \AA\ and 0.09, respectively. Together with our NpH results,
these comparisons support the general notion that structural properties are well 
converged in both simulation schemes.

\begin{table}
\caption{Distribution of water molecules with a given number of hydrogen bonds.
         The percentages correspond to the height of each bar in the histograms
         shown in Fig.\ \ref{fig:hbonds}. The largest statistical error is found
         for the fourfold hydrogen-bonded molecules and amounts to 1.1\% and 1.7\%
         for {\sc Q}uantum-{\sc Espresso} and {\tt CP2K}, respectively.         
         The average number of hydrogen bonds per water molecule is given in the
         last column and the experimental estimate from Ref.\ \citenum{soper_JCP1997}
         is reported for comparison. The statistical error on the average number
         of hydrogen bonds is less than 0.01. The statistical errors are estimated by
         treating nonoverlapping trajectories of the duration of 2.5 ps as indepedent.}
\begin{ruledtabular}
\begin{tabular}{lcccccc}
   & \multicolumn{6}{c}{Number of hydrogen bonds} \\
   \cline{2-7}
                                &  1   &  2   &  3   &  4   &  5   & average \\
   \hline
   {\sc QE}                     &  1\% &  8\% & 29\% & 56\% &  6\% & 3.58 \\
   {\tt CP2K}                   &  2\% & 10\% & 31\% & 51\% &  6\% & 3.49 \\
   Expt.(Ref.\ \citenum{soper_JCP1997})          & --   & --   & --   & --   & --   & 3.58 
\end{tabular}
\end{ruledtabular}
\label{tab:hbonds}
\end{table}

\begin{figure}
\centering
\includegraphics[width=8.5cm]{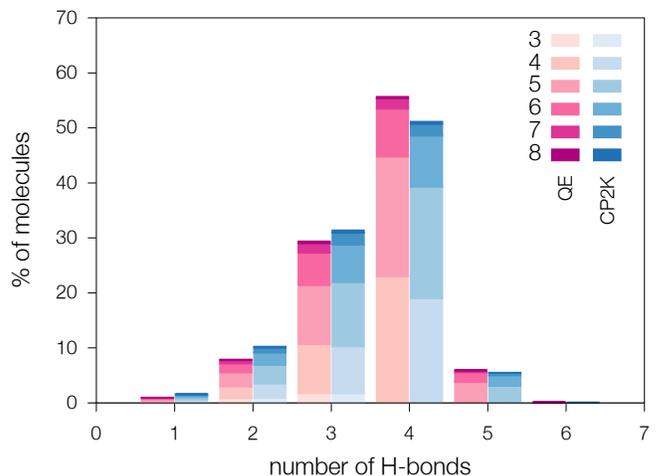}%
\caption{Distributions of water molecules with a given number of hydrogen bonds 
         as obtained with  {\sc Q}uantum-{\sc Espresso} (QE) and {\tt CP2K} 
         (cf.\ also Table\ \ref{tab:hbonds}).
         The finer subdivisions within each bar, here indicated with a gradient of
         colors, show the associated distribution of the total coordination number.
         Both the total coordination number and the number of hydrogen bonds are
         defined using the same cutoff distance of 3.5 \AA. For this reason, the
         former cannot be smaller than the latter.}
 	\label{fig:hbonds}
\end{figure}

The anomalous behavior of liquid water has generally been connected to the complex 
bonding network formed by the water molecules.\cite{mills_JPC1973,krynicki_Farad1978} 
The water network changes with the breaking and the formation of hydrogen bonds 
as regulated by the considered thermodynamic conditions. A first approximate description
of these complex molecular networks can be given by the average number of hydrogen
bonds. This quantity is not directly accessible to experiments. However, a value
of 3.58 per molecule has been inferred from experimental data for water at ambient
conditions.\cite{soper_JCP1997} We here define the hydrogen bond using a purely 
geometrical criterion commonly used in the literature.\cite{luzar_PRL1996,%
soper_JCP1997,schwegler_PRL2000,miceli_JCP2015} We consider two water molecules 
to be hydrogen-bonded when their oxygen-oxygen distance is at most 3.5 \AA\ and 
their hydrogen-bond angle $\angle$OHO is simultaneously larger than 140$^{\circ}$.
Based on this criterion the calculated average number of hydrogen bonds is 
$(3.58 \pm 0.007)$ as obtained with {\sc Q}uantum-{\sc Espresso} and 
$(3.49 \pm 0.010)$ with {\tt CP2K}, where the uncertainties are estimated by 
treating nonoverlapping trajectories of the duration of 2.5 ps as indepedent. 
The difference between the two simulations exceeds the statistical 
error estimated in this way, but remains small. Similar values
for the average number of hydrogen bonds per molecule have been found through 
NVE-MD simulations based on other semilocal density functionals.\cite{todorova_JPCB2006,%
schwegler_PRL2000} In Table \ref{tab:hbonds}, we report the percentage of molecules
with a given number of hydrogen bonds. The majority percentage associated to the
fourfold hydrogen-bonded molecules is subject to the largest statistical error, 
which amounts to $\approx$1.1\% and $\approx$1.7\% for {\sc Q}uantum-{\sc Espresso} 
and {\tt CP2K}, respectively. From the comparison of the results obtained with 
the two codes, one notices that {\tt CP2K} gives a slightly lower number of hydrogen
bonds when compared to {\sc Q}uantum-{\sc Espresso}. In fact, water molecules 
with less than four hydrogen bonds are found to be more likely.

To investigate the local order and the overall structural organization at a higher
level detail, we proceed with a finer analysis of the short-range order by calculating
the total coordination number of a molecule with a given number of hydrogen bonds. 
In practice, let us consider the most probable situation in which a molecule shows
four hydrogen bonds. Within the cutoff distance of 3.5 \AA\ that defines the first
coordination shell, the molecule might show a solvation shell containing more than
four water molecules. These results are illustrated in Fig.\ \ref{fig:hbonds}. The
heights of the histogram bars correspond to the percentages reported in Table 
\ref{tab:hbonds}, while the gradient of colors within the same bar indicates the
percentages of molecules with higher coordination number. We have shown that this
further analysis is very sensitive to the adopted theoretical scheme.\cite{miceli_JCP2015}
In particular, we have demonstrated that even though rVV10-b6.3 and rVV10-b9.3 
yield very close average numbers of hydrogen bonds (3.55 and 3.59, respectively)
the finer distribution illustrating total coordination numbers differs noticeably.
In Fig.\ \ref{fig:hbonds}, we show that the two codes produce close results for 
the hydrogen-bond network also at this finer level of detail.


\begin{figure}
\centering
\includegraphics[width=8.5cm]{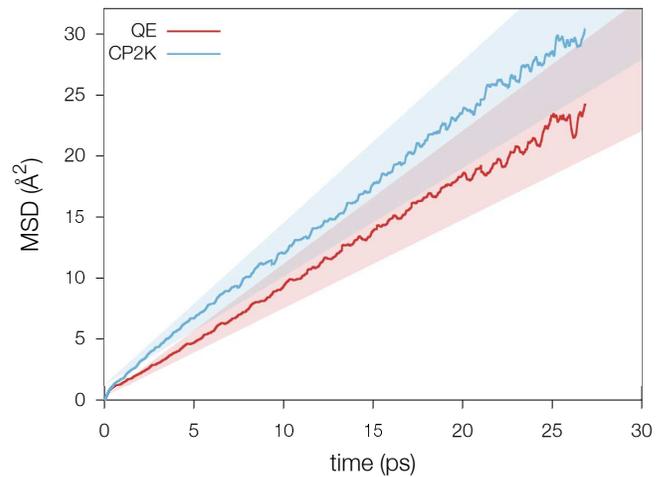}%
\caption{Mean square displacement as a function of time for the NVE-MD runs
         obtained with {\sc Q}uantum-{\sc Espresso} (QE) and {\tt CP2K}. The 
         shaded areas represent an estimation of the statistical errors in 
         evaluating the MSD.}
 	\label{fig:msd}
\end{figure}

\begin{table}
\caption{Self-diffusion coefficient $D^\textrm{sim}$ of liquid water obtained 
         from our NVE-MD runs obtained with {\sc Q}uantum-{\sc Espresso} (QE) and {\tt CP2K}. 
         The calculated values are compared to reference values $D^\textrm{ref}$, which 
         represent the experimental diffusion coefficient modified to account for the 
         finite-size effect pertaining to the cell size used in the simulations. 
         For reference, we also report the actual experimental data extracted 
         from Ref.\ \citenum{holz_PCCP2000}.}
\begin{ruledtabular}
\begin{tabular}{lccc}
                       & $T$ (K) &  $D^\textrm{sim}$ (cm$^2$/s) & $D^\textrm{ref}$  (cm$^2$/s)  \\
  \hline
  {\sc QE}             & 350     &   $(1.50\pm0.30)  \cdot 10^{-5}$       &  $4.3 \cdot 10^{-5}$ \\
  {\tt CP2K}           & 350     &   $(1.80\pm0.36)  \cdot 10^{-5}$       &  $4.3 \cdot 10^{-5}$ \\
  \hline
  &&& $D^\textrm{expt.}$  (cm$^2$/s)  \\
  \hline
  Expt.(extrapolated)  & 350 &      & $6.2 \cdot 10^{-5}$ \\
  Expt.                & 300 &      & $2.4 \cdot 10^{-5}$ 
 \end{tabular}
 \end{ruledtabular}
\label{tab:msd}
\end{table}

We calculate the self-diffusion coefficient of liquid water using Einstein's 
relation. The mean square displacement (MSD) as a function of time results from 
averaging over all water molecules. We average over the NVE-MD trajectories 
choosing initial times separated by 2 ps. In Fig.\ \ref{fig:msd}, we report the 
MSDs as a function of time and the respective diffusion coefficients with their 
relative statistical errors are summarized in Table\ \ref{tab:msd}. 
We estimate the statistical errors on the diffusion coefficients by performing a 
large set of independent molecular dynamics simulations based on an empirical 
force field. For this, we use simulations with the same set-up (duration, supercell 
size, thermodynamic conditions) as for the {\it ab initio} MD. The resulting 
percent error on the self-diffusion coefficient is then assumed to apply to 
the {\it ab initio} MD. Within the statistical errors determined in this way 
the two codes give consistent diffusion coefficients. For comparison, in an earlier 
comparison of the same kind, Kuo {\it et al.}\ found
diffusion coefficients with relative differences ranging between 25\%\ and 50\%\
of the plane-wave result.\cite{kuo_JPCB2004} At variance, the comparison with the
complete-basis-set simulations achieved with a discrete variable representation 
\cite{lee_JCP2006,lee_JCP2007} cannot directly be compared with our results. Indeed,
the latter simulations have been carried out with the Car-Parrinello method, which
leads to biased diffusion coefficients.\cite{tangney_JCP2002,grossman_JCP2003,kuo_JPCB2004}

The comparison of the calculated values with experiments requires some care. In 
fact, it is well known that dynamical properties suffer from finite-size effects
more than structural properties. Generally, this leads to an underestimation of 
the self-diffusion coefficient. To account for the finite-size effect, we compare
the calculated diffusion coefficients $D^\textrm{sim}$ with reference values 
$D^\textrm{ref}$, derived from experimental data. Here, $D^\textrm{ref}$ represents
the experimental self-diffusion coefficient at the temperature of 350 K, modified
to account for the cell size used in the simulation.\cite{miceli_JCP2015} 
We notice that, although the rVV10-b9.3 functional improves the description of 
the dynamical properties of liquid water, the experimental value is still
underestimated by about a factor 2.5 at this level of theory.

 
\begin{figure}
\centering
\includegraphics[width=8.5cm]{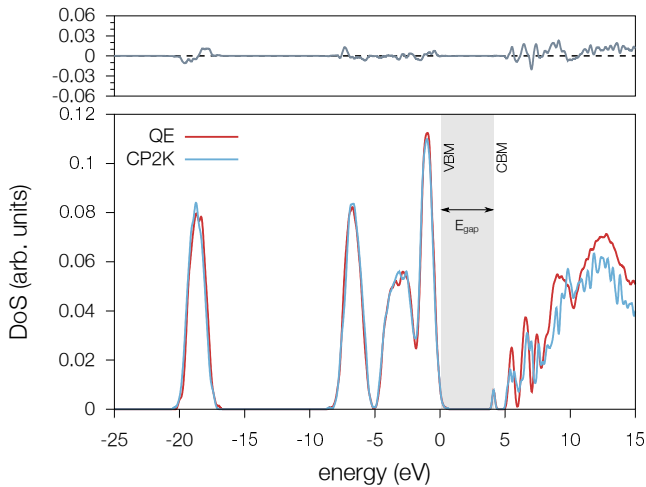}%
\caption{Electronic density of states averaged over 20 configurations of liquid water,
         separated by 50 fs, taken from the tail of the NVE-MD trajectories.
         The electronic structures calculated with the two codes are aligned through
         the valence band edge and their difference is plotted in the upper part 
         of the panel. The shaded region represents the band gap.}
 	\label{fig:dos}
\end{figure}

\begin{table}
\caption{Energy-level separations obtained by averaging over the {\sc Q}uantum-{\sc Espresso} (QE)
         and {\tt CP2K} MD trajectories. $\varepsilon_{\rm c}$, $\varepsilon_{\rm v}$, and 
         $\varepsilon_{{\rm O}2s}$ refer to the conduction band edge, the valence band edge
         and the average level of O 2$s$, respectively. The statistical
         errors are given in parentheses. Energies are given in eV.}
\begin{ruledtabular}
\begin{tabular}{lccc}
                            & $\varepsilon_{\rm v}-\varepsilon_{{\rm O}2s}$ &  
                              $\varepsilon_{\rm c}-\varepsilon_{{\rm O}2s}$ & 
                              $\varepsilon_{\rm c}-\varepsilon_{\rm v}$   \\
  \hline
  {\sc QE}                  & 18.61 (0.016) & 22.71 (0.004)  & 4.10 (0.016) \\
  {\tt CP2K}                & 18.72 (0.016) & 22.92 (0.005)  & 4.20 (0.016) \\
 \end{tabular}
 \end{ruledtabular}
\label{tab:edges}
\end{table}

Next, we focus on the electronic structure of liquid water. Describing its 
electronic properties does not only have a fundamental interest, but is also 
critical for applications involving the water splitting process. 
In Fig.\ \ref{fig:dos}, we compare the electronic density of states  
of liquid water as obtained with  {\sc Q}uantum-{\sc Espresso} and {\tt CP2K}. 
These have been obtained via a statistical average over 20 different water 
configurations regularly spaced by 50 fs, which have been extracted from the 
NVE-MD trajectories at 350 K.
The two electronic structures are superimposed and aligned through the valence 
band edge. The results from the two codes are in excellent agreement. As 
one can see from the difference of the two curves in the upper part of 
Fig.\ \ref{fig:dos}, imperceptible differences are present for occupied states. 
The two curves slightly depart from each other for the unoccupied states 
at higher energies. The latter behavior should be ascribed to the use of 
different basis sets for the expansion of the electronic wave functions in 
the two codes.

From a theoretical point of view, the definition of band gap for a disordered insulator 
might lead to ambiguities and be subject to finite size effects.\cite{ambrosio_JCP2015} 
Nevertheless, these issues do not apply here because we are interested in comparing
two codes for simulations with identical supercells.
The band gap is thus determined as the energy difference between the lowest unoccupied 
(LUMO) and the highest occupied (HOMO) molecular orbitals over the molecular dynamics 
trajectory. In Table\ \ref{tab:edges}, we report the calculated values for the band gap 
($\varepsilon_{\rm c}-\varepsilon_{\rm v}$) and for the conduction and valence band-edge 
positions with respect to the average O 2$s$ level ($\varepsilon_{\rm v}-\varepsilon_{{\rm O}2s}$
and $\varepsilon_{\rm c}-\varepsilon_{{\rm O}2s}$, respectively). The determined band 
gaps are 4.10 and 4.20 eV as obtained with {\sc Q}uantum-{\sc Espresso} and {\tt CP2K}, 
respectively. The energy separations $\varepsilon_{\rm v}-\varepsilon_{{\rm O}2s}$  
similarly agree within 0.1 eV. For $\varepsilon_{\rm c}-\varepsilon_{{\rm O}2s}$, 
the difference is only slightly larger (0.2 eV).  Overall, This close accord for 
the electronic levels further supports the agreement between the two codes.

\section{Conclusions}
\label{sec:conclusions}
In this work, we compare results obtained through NpH and NVE molecular dynamics 
simulations performed using plane-wave and atomic-orbital basis sets, as implemented in 
{\sc Q}uantum-{\sc Espresso} and {\tt CP2K}, respectively. In both frameworks, 
we used the same nonlocal density functional approximation accounting for van 
der Waals interactions. Noticeably, the schemes based on plane waves and atomic 
orbitals yield results for structural, dynamical, and electronic properties 
in overall very good agreement with each other.  Hence, the quality of this 
agreement allows one to envisage equivalently the use of either implementation 
in the study of liquid water or aqueous solutions.

\acknowledgments
The authors thank S. Goedecker for fruitful interactions. This work has been 
performed in the context of the National Center of Competence in Research (NCCR)
``Materials' Revolution: Computational Design and Discovery of Novel Materials
(MARVEL)'' of the Swiss National Science Foundation. We used computational resources
of CSCS and CSEA-EPFL.

\end{document}